\documentclass{article}

\usepackage[final,nonatbib]{nips_2016}

\usepackage[utf8]{inputenc} 
\usepackage[T1]{fontenc}    
\usepackage{url}            
\usepackage{booktabs}       
\usepackage{amsfonts}       
\usepackage{nicefrac}       
\usepackage{microtype}      
\usepackage{algorithm}
\usepackage{algorithmic}
\usepackage{amsbsy,amssymb,amsmath,amsthm,cases,mathrsfs}
\usepackage{float}
\usepackage{color}
\usepackage{wrapfig}
\usepackage{paralist}

\let\oldbibliography\thebibliography
\renewcommand{\thebibliography}[1]{\oldbibliography{#1}
\setlength{\itemsep}{0pt}}
\theoremstyle{plain}
\newtheorem{theorem}{Theorem}[section]
\newtheorem{lemma}[theorem]{Lemma}

\theoremstyle{definition}
\newtheorem{definition}[theorem]{Definition}

\theoremstyle{remark}

\newcommand{\R}{\mathbb{R}}

\newcommand{\PP}{\mathcal{P}}

\newcommand{\suchthat}{\;:\;}
\newcommand{\abs}[1]{\left|#1\right|}
\newcommand{\norm}[1]{\left\|#1\right\|}
\newcommand{\size}[1]{\left|#1\right|}
\newcommand{\linspan}[1]{\operatorname{span}\left(#1\right)}
\newcommand{\prob}[1]{\operatorname{Pr}\left(#1\right)}
\usepackage{relsize}
\usepackage{graphicx} 
\usepackage{titlesec}
\usepackage{caption}
\usepackage{subcaption}
\title{On Sampling and Greedy MAP Inference of Constrained Determinantal Point Processes}

\author{
  Tarun Kathuria, Amit Deshpande \\
  Microsoft Research, India\\
  \texttt{t-takat@microsoft.com, amitdesh@microsoft.com}
}

\begin{document}
\setlength{\abovedisplayskip}{1em}
\setlength{\belowdisplayskip}{1em}
\maketitle

\begin{abstract}
Subset selection problems ask for a small, diverse yet representative subset of the given data. When pairwise similarities are captured by a kernel, the determinants of submatrices provide a measure of diversity or independence of items within a subset. Matroid theory gives another notion of independence, thus giving rise to optimization and sampling questions about Determinantal Point Processes (DPPs) under matroid constraints. Partition constraints, as a special case, arise naturally when incorporating additional labeling or clustering information, besides the kernel, in DPPs. Finding the maximum determinant submatrix under matroid constraints on its row/column indices has been previously studied. However, the corresponding question of sampling from DPPs under matroid constraints has been unresolved, beyond the simple cardinality constrained $k$-DPPs.

We give the first polynomial time algorithm to sample exactly from DPPs under partition constraints, for any constant number of partitions. We complement this by a complexity theoretic barrier that rules out such a result under general matroid constraints. Our experiments indicate that partition-constrained DPPs offer more flexibility and more diversity than $k$-DPPs and their naive extensions, while being reasonably efficient in running time.

We also show that a simple greedy initialization followed by local search gives improved approximation guarantees for the problem of MAP inference from $k$-DPPs on \emph{well-conditioned} kernels. Our experiments show that this improvement is significant for larger values of $k$, supporting our theoretical result.
\end{abstract}

\section{Introduction}
\label{introduction}
Selecting a small, diverse yet representative subset of the given data is an important problem underlying feature/exemplar selection in machine learning, sensor placements \cite{Krause2008}, row/column subset selection in linear algebra \cite{Deshpande2010}, and coresets in computational geometry \cite{Clarkson2010}. Subset selection from a given \emph{ground set} $\mathcal{Y}$ of $m$ items is often formulated as an optimization or a sampling problem over all the $2^{m}$ possible subsets, with constraints on the size and the diversity of the desired subset. If an $m$-by-$m$ positive semidefinite kernel matrix $K$ captures all pairwise similarities, then matrix determinants and log-determinants help formulate diversity in the form of volume \cite{Deshpande2010,Deshpande2006,GZT1997}, entropy \cite{Krause2008}, and repulsion of Fermions \cite{Shirai2003}. Geometrically, this corresponds to picking a subset of vectors that are long and as linearly independent as possible. Matroid theory provides another compelling notion of independence in subsets that underlies spanning trees, matchings etc. Thus, a large class of subset selection problems can be formulated as either optimization or sampling problems, stated using determinants or log-determinants, and subject to various matroid constraints.

Motivated by the above, we model the diversity of a subset $S \subseteq [m]$ by the determinant of its corresponding $\size{S}$-by-$\size{S}$ submatrix $K_{S, S}$ in a given $m$-by-$m$ positive semidefinite kernel $K$, and summarize the previous work on optimization and sampling questions to lay the premise.

The optimization problem for subset selection corresponds to maximizing $\det(K_{S, S})$, or equivalently $\log \det(K_{S, S})$, subject to additional cardinality or matroid constraints. This is NP-hard even in the absence of constraints \cite{Civril2009}. However, the submodularity of $f(S) = \log\det(K_{S,S})$ gives a polynomial time $1/4$-approximation for maximizing $\log \det(K_{S, S})$ \cite{GillenwaterMAP2013}. When $\log\det(K_{S,S})$ is monotone (e.g., when the smallest eigenvalue $\lambda_{\min}(K) \geq 1$), a greedy algorithm gives $(1 - 1/e)$-approximation, even under cardinality or matroid constraints \cite{Nemhauser1978}. However, multiplicative approximations for maximizing $\log \det(K_{S,S})$ do not imply similar results for $\det(K_{S,S})$. The best known polynomial time algorithm for maximizing $\det(K_{S,S})$ over subsets of size $k$ gives an approximation guarantee of $1/e^{k+o(k)}$ \cite{Nikolov15}, and it is NP-hard to do better than $1/c^k$, for some $c > 1$ \cite{SummaEFM15}. Recently, the same approximation guarantee was achieved for maximizing $\det(K_{S,S})$ under partition constraints on $S$, which is a special case of matroid constraints, via a \emph{geometric concave program} \cite{NikolovSingh16}.

On the other hand, the sampling problem for subset selection corresponds to sampling a subset $S \subseteq [m]$ with probability proportional to $\det(K_{S,S})$. These distributions are known as Fermion Point Processes or Determinantal Point Processes (DPPs), with remarkable applications in probability, statistical physics, and random matrix theory \cite{Borodin2009,BorodinRains2005,Lyons2003,Hough2006}. Sampling exactly from a DPP and its cardinality-constrained variant $k$-DPP can both be done in polynomial time \cite{Deshpande2010,Hough2006}. This has found applications in document summarization \cite{Lin2012}, object retrieval \cite{Affandi2014}, sensor placement \cite{Krause2008}. We point the reader to \cite{Kulesza2011,Kulesza2012} for efficient algorithmic sampling, inference, and computation of the partition functions, marginals, conditional probabilities etc. of DPPs and $k$-DPPs, and their numerous applications in machine learning.

Beyond the simple cardinality-constrained $k$-DPPs, previous work does not provide efficient algorithms for exact sampling from a DPP under general matroid constraints. The difficulty arises from the computation of its partition function, namely, the sum of $\det(K_{S,S})$ over subsets $S$ that satisfy the given matroid constraints. For general matroid constraints, this sum does not have a nice, closed form expression (unlike in the case of DPPs and $k$-DPPs).

The main purpose of our paper is to study sampling from DPPs under general matroid constraints. We give the first polynomial time algorithm for exact sampling from DPPs under partition constraints, for any constant number of partitions. We complement this by a complexity theoretic barrier that rules out such a result under general matroid constraints. To be precise, DPPs under transversal matroid constraints can simulate random perfect matchings in a bipartite graph, whose partition function is $\#P$-hard (refer to Appendix 1 for details).

Partition constraints arise naturally when we have some inherent labeling or clustering of the items that is not captured by the DPP kernel. In such cases, DPPs may not capture the diversity correctly, leading to over/under-representation as explained below.
\begin{enumerate}
  \item Google image search for the query ``jaguar'' fetches images of the animal, the cars from the company ``Jaguar'', the logo of the car company etc. While the search engine recognizes these categories, the top results are dominated by the animal and the cars, causing the company logo to be under-represented. $k$-DPPs based only on image kernels are unlikely to fix this problem as the stylistic features of the logo are similar to that of the animal.

  \item Consider a facial image database of people with different expressions and lighting conditions. Kernels based on SIFT and dot-products do not capture lighting conditions well but are still desirable due to their computational efficiency. Now if we want a subset of facial images where the people as well as the lighting conditions therein are distinct, or if want certain lighting conditions more prominently in our subset, then the kernel alone may not be able to capture such constraints.
  \end{enumerate}

In these two examples, partition constraints arise naturally from the image categories and the lighting conditions, respectively. They could also arise via clustering based on features not captured by the kernel. Thus, when the ground set has a disjoint partition $\mathcal{Y} = \PP_{1} \biguplus \PP_{2} \biguplus \dotsc \biguplus \PP_{p}$ given apriori, and we want to pick a subset $S$ with $k_{1}$ items from $\PP_{1}$, $k_{2}$ items from $\PP_{2}$, and so on, we define a Partition-DPP as the corresponding conditioning of DPP under these constraints on $S$.

It is imperative to contrast Partition-DPP with two natural extensions of $k$-DPP, namely, \begin{inparaenum}[(a)] \item sampling $k = k_{1} + \dotsc + k_{p}$ images from the entire set by $k$-DPP and \item sampling from $p$ independent $k_{i}$-DPPs each on part $\PP_{i}$, respectively. \end{inparaenum} $k$-DPP suffers from over/under-representation and is not guaranteed to pick $k_{i}$ items from $\PP_{i}$, due to interference between inter-partition and intra-partition diversity. In the example of different people under different lighting conditions, if our partitions are given by the lighting conditions and we sample from independent $k_{i}$-DPPs on each part, we may end up picking the same person from different lighting conditions in our subset. It is also easy to construct counterexamples to rule out $k$-DPP followed by rejection sampling based on the partition constraints. 

Our algorithm for Partition-DPPs is based on a multivariate generalization of the characteristic polynomial of a matrix that may be of independent interest. Since the effectiveness of DPPs in modeling diversity has been demonstrated in \cite{Kulesza2011,Kulesza2012}, our experiments focus on comparing our algorithm against the two natural extensions of $k$-DPP stated above, and highlight the over/under-representation, or equivalently, the violation of partition constraints.

In addition, we quantitatively improve the MAP inference approximation guarantee for $k$-DPPs, when the DPP kernel is \emph{well-conditioned}, and show that a simple greedy algorithm followed by local search provides almost as good an approximation guarantee for maximizing $\det(K_{S, S})$ over $k$-sized subsets as the expensive convex program in \cite{Nikolov15}. Our experiments show that this improvement is significant for larger values of $k$, supporting our theoretical result.


\section{Setup and basic definitions}
Let $[m]$ denote the set $\{1, 2, \dotsc, m\}$ used to index the $m$ items in our ground set $\mathcal{Y}$. We identify $\mathcal{Y}$ with $[m]$ and may use them interchangeably. For any positive semidefinite matrix $K \in \R^{m \times m}$ and a subset $S \subseteq [m]$, we use $K_{S, S}$ to denote the corresponding $\size{S}$-by-$\size{S}$ submatrix of $K$ whose both the row and the column indices are in $S$. For a rectangular matrix $A \in \R^{m \times n}$, we denote its rows by $a_{1}, a_{2}, \dotsc, a_{m} \in \R^{n}$, and we use $A_{S}$ to denote the $\size{S}$-by-$n$ row-submatrix formed by row indices in $S$. Let $\linspan{S}$ be the linear subspace spanned by the rows of $A_{S}$, and $\pi_{S}(A)$ be the orthogonal projection onto $\linspan{S}$.

For $K = AA^{T}$, we have $K_{S, S} = A_{S} A_{S}^{T}$. $k$-DPP defines a distribution over subsets $S \subseteq [m]$ of size $k$, where the probability of picking $S$ is proportional to $\det(K_{S, S}) = \det(A_{S} A_{S}^{T})$, which is also proportional to the squared volume of the parallelepiped formed by the rows of $A_{S}$.

\begin{definition}{($k$-DPP)} Given a positive semidefinite matrix $K \in \R^{m \times m}$ as DPP kernel, the cardinality-constrained DPP $k$-DPP is defined by the following distribution on subsets $B \subseteq [m]$ of size $k$.
\[
\prob{B} = \frac{\det(K_{B,B})}{\sum\limits_{S \suchthat |S|=k} \det (K_{S, S})} = \frac{\displaystyle \det(A_{B} A_{B}^{T})}{\sum\limits_{S \suchthat \size{S}=k} \det (A_S A_S^T)}.
\]
\end{definition}

The following linear algebraic identity about the coefficients of the characteristic polynomial is at the heart of the polynomial time sampling algorithm for $k$-DPPs in \cite{Deshpande2010,Kulesza2011} (see Algorithm \ref{alg:kdppalgo}).
\begin{theorem} (Proposition 3.2 in \cite{Deshpande2006}) For any $A \in \R^{m \times n}$, let the characteristic polynomial of $A A^T \in \R^{m\times m}$ be $\det(A A^{T} - xI) = x^m + c_{m-1}x^{m-1} + \ldots + c_{0}$. Then
\[
\sum\limits_{S \suchthat \size{S}=k} \det(A_{S} A_{S}^{T}) = \abs{c_{m-k}}, \quad \text{for~ $1 \leq k \leq m$}.
\]
\end{theorem}
We refer the reader to \cite{Deshpande2006,Martin} for the proof of the above theorem. 

\begin{algorithm}[tb]
   \caption{Sampling from $k$-DPP (Algorithm 1 in \cite{Deshpande2010})}
   \label{alg:kdppalgo}
\begin{algorithmic}
   \STATE {\bfseries Input:} Matrix $A \in \R^{m \times n}$ and $1 \leq k \leq $ rank(A)
   \STATE {\bfseries Output:} Subset $S$ of $k$ rows of $A$ picked with probability proportional to $\det(A_S A_S^T)$
   \STATE Initialize $S \leftarrow \emptyset$ and $B \leftarrow A$.
   \FOR{$t=1$ {\bfseries to} $k$}
   \FOR{$i=1$ {\bfseries to} $m$}
   \STATE Compute $p_i = \norm{b_{i}}^{2} \abs{c_{m-k+t}(C_{i} C_{i}^{T})}$,
   \\ where $C_i$ is the matrix obtained by projecting each row of $B$ orthogonal to $b_i$.
   \ENDFOR

   \STATE Pick $i$ with probability proportional to $p_i$.
   \STATE $S \leftarrow S\cup\{i\}$ and $B\leftarrow C_i$
   \ENDFOR
   \STATE \textbf{return} $S$
\end{algorithmic}
\end{algorithm}

Given a partition of the ground set into disjoint parts, a partition constraint on $S$ says that it must have $k_{1}$ items from the first part, $k_{2}$ from the second part, and so on.
\begin{definition}{(Partition Constraint)} Given a disjoint partition $\PP$ as $\mathcal{Y} = [m] = \PP_{1} \biguplus \PP_{2} \biguplus \dotsb \biguplus \PP_{p}$ into $p$ parts of sizes $m_{1}, m_{2}, \dotsc, m_{p}$, respectively, a partition constraint on a subset $S \subseteq [m]$ is defined by a $p$-tuple $(k_{1}, k_{2}, \dotsc, k_{p})$, and $S$ is said to satisfy this partition constraint if $S$ has $k_{i}$ elements from part $\PP_{i}$, respectively, i.e., $\size{S \cap \PP_{i}} = k_{i}$, for $1 \leq i \leq p$.
\end{definition}

Now we define a multivariate characteristic polynomial of $AA^{T}$ with respect to the partition $\mathcal{Y} = [m] = \PP_{1} \biguplus \PP_{2} \biguplus \dotsb \biguplus \PP_{p}$, by splitting $xI$ in $\det(AA^{T} - xI)$ into smaller identity matrices with different variables, one for each partition.
\begin{definition}{(Multivariate Characteristic Polynomial)} \label{multiCharPolyDef} For any $A \in \R^{m \times n}$ and a partition $\PP$ as $\mathcal{Y} = [m] = \PP_{1} \biguplus \PP_{2} \biguplus \dotsb \biguplus \PP_{p}$, we define the corresponding multivariate characteristic polynomial of $AA^{T}$ as
\[
\det(A A^T - x_1 I_1 - \ldots - x_p I_p) = \mathlarger{\mathlarger{\sum}}_{i_p=0}^{m_p} \ldots \mathlarger{\mathlarger{\sum}}_{i_1=0}^{m_1} c_{i_1, \dotsc, i_p} x_1^{i_1}\ldots x_p^{i_p},~ \text{where}~ (I_{t})_{i,i} = \begin{cases} 1 &  \text{if $i \in \PP_{t}$} \\ 0 & \text{otherwise} \end{cases}.
\]
\end{definition}

\section{Main Results}
In this section, we first present our sampling algorithm for Partition-DPPs that uses the multivariate characteristic polynomial to generalize $k$-DPP sampling. In the later subsection, we present our results about quantitative improvements for MAP inference in $k$-DPPs, for \emph{well-conditioned} DPP kernels, using greedy algorithm followed by local search.
\subsection{Partition-DPPs}
As argued earlier, $k$-DPPs and independent $k_{i}$-DPPs for each partition cannot be used for exact sampling from Partition-DPPs. It is also easy to see that $k$-DPPs followed by rejection sampling based on the partition constraints cannot give an efficient algorithm either. In \cite{IyerSPP15} (Section 3.6), the authors consider Submodular Point Processes and Log-submodular Point Processes (of which DPPs are a special case) under partition constraints of the type $\size{S \cap \PP_i} \leq k_{l}$, for $1 \leq l \leq p$. They argue that the partition function $Z$, that is, the sum of $\det(K_{S, S})$ over subsets $S$ satisfying such partition constraints, splits as $Z = \sum_{l=1}^{p} \sum_{k=1}^{k_l} Z^{k}_{\PP_{l}}$, where each $Z^{k}_{\PP_l}$ is the partition function of a $k$-DPP defined only on the elements of $\PP_l$. This partition function $Z$ actually corresponds a mixture of cardinality-constrained DPPs on each part, and does not apply to Partition-DPP as in our definition \cite{BilmesPersonalComm}. We explain this with an example in the appendix. 

Now we show how the coefficients of the multivariate characteristic polynomial are related to the partition functions for Partition-DPPs, that is, the sum of $\det(K_{S, S})$ over all subsets $S$ that satisfy a given partition constraint.
\begin{lemma}\label{multicoeff} Consider the multivariate characteristic polynomial of $A A^T \in \R^{m \times m}$ with respect to partition $\PP$ given by $\mathcal{Y} = \PP_{1} \biguplus \PP_{2} \biguplus \dotsb \biguplus \PP_{p}$, as in Definition \ref{multiCharPolyDef}. Then
\[
\sum\limits_{S \suchthat \size{S \cap \PP_{i}} = k_{i},~ \text{for}~ 1 \leq i \leq p} \det(A_S A_S^T) = \abs{c_{m_{1} - k_{1}, \dotsc, m_{p} - k_{p}}}, \qquad \text{for any}~ 0 \leq k_{i} \leq m_{i}.
\]
\end{lemma}
\begin{proof}(\textit{Sketch.}) We first expand $\det(AA^{T} - x_{1} I_{1} - \dotsc - x_{p} I_{p})$ using the Leibniz formula for determinant \cite{LeibnizWiki}, and then collect the terms for each monomial together. Again by Leibniz formula, the collected terms sum up $\det(A_{S} A_{S}^{T})$ over certain subsets $S$, which happen to be the ones that satisfy the corresponding partition constraints.
\end{proof}

As in the case of $k$-DPPs or volume sampling \cite{Deshpande2010}, it helps to think of Partition-DPP as a distribution over ordered tuples rather than sets. Specifically, $\prob{X_1 = i_1, \ldots, X_k = i_k}$ is equal to
\[
\begin{cases}\dfrac{\det(A_{\{i_1, \dotsc,i_k\}} A_{\{i_1, \dotsc, i_k\}}^T)}{k! \sum\limits_{S \suchthat \size{S \cap \PP_{i}} = k_{i},~ \text{for}~ 1 \leq i \leq p} \det(A_S A_S^T)}, & \qquad \substack{\displaystyle \text{for distinct $i_1, \dotsc, i_k$ that} \\ \displaystyle \text{satisfy partition constraints $\PP$}} \\ 0, & \qquad \text{otherwise} \end{cases}
\]
Then the marginal probabilities $\prob{X_{t+1} = i~ |~ X_{1} = i_{1}, \dotsc, X_{t} = i_{t}}$ have the following interpretation in terms of the co-efficients of certain derived multivariate characteristic polynomials. This generalizes a corresponding theorem for $k$-DPP in \cite{Deshpande2010} and gives the technical core of our algorithm.
\begin{theorem} \label{margProbLemma} Let $(i_1, \dotsc, i_t)\in [m]^t$ be such that $\prob{X_1 = i_1, \dotsc, X_{t}=i_{t}} > 0$, in the above Partition-DPP extended to $k$-tuples. Let $S =\{i_1, \dotsc, i_t \}$ and $\size{S \cap \PP_l} = t_l$, for  all $l \in [p]$, with $t = \sum\limits_{l=1}^{p} t_l$. For any $i \in \PP_j$, let $B= A - \pi_S(A)$ and $C_i = B - \pi_{\{i\}}(B) = A - \pi_{S\cup\{i\}}(A)$. Then
\begin{align*}
\prob{X_{t+1} = i~ |~ X_{1} = i_1, \dotsc, X_t = i_t} & = \frac{\norm{b_{i}}^{2} \abs{c'(C_i C_i^T)}}{(k-t)\abs{c''(B B^T)}}, \quad \text{where} \\ 
c'(C_i C_i^T) & = c_{m_1-k_1+t_1, \dotsc, m_j-k_j+t_j+1, \dotsc, m_p-k_p+t_p}(C_i C_i^T) \\ 
c''(B B^T) & = c_{m_1-k_1+t_1, \dotsc, m_j-k_j+t_j, \dotsc, m_p-k_p+t_p}(B B^T).
\end{align*}
\end{theorem}
\begin{proof}(\textit{Sketch.}) Being conditional probability, $\prob{X_{t+1} = i~ |~ X_{1} = i_{1}, \dotsc, X_{t} = i_{t}}$ can be written as $\prob{X_1 = i_1, \dotsc, X_t = i_t, X_{t+1}= i}/\prob{X_1 = i_1, \dotsc, X_t = i_t}$, where the numerator and the denominator sum up probabilities over the indices in $(i_1, \dotsc, i_k)$ not fixed by the marginal probability expression. Thus, the numerator sums over subsets of the remaining indices that satisfy partition constraints of the form, $k_1 - t_1, \dotsc, k_j - (t_j + 1), \dotsc, k_p - t_p$. The denominator is a similar sum over subsets with partition constraints of the form $k_1 - t_1, \dotsc, k_j - t_j, \dotsc, k_p - t_p$. Some further manipulations of the subdeterminants in the sums and Lemma \ref{multicoeff} complete the proof.
\end{proof}

With this theorem in hand, Algorithm \ref{alg:partdppalgo} follows naturally. 
By computing the co-efficients of multivariate characteristic polynomials as in Definition \ref{multiCharPolyDef}, we can do exact sampling from DPPs under partition constraints. Algorithm \ref{alg:partdppalgo} runs in time $\mathcal{O}\left(m^{p+4} nk/p^p\right)$, where $k = \sum_{l=1}^p k_l$. One way to compute the coefficients of a univariate characteristic polynomial is by evaluating the polynomial $\det(A-xI)$ at $m+1$ distinct points $x$, which gives a system of linear equations in its coefficients that can be solved by inverting a Vandermonde matrix. This generalizes to the multivariate case, albeit inefficiently, because it can have $\prod_{l=1}^{p} (m_i + 1) \leq (1 + m/p)^p$ coefficients (by AM-GM inequality). Also this interpolation approach computes all the coefficients, when the algorithm requires only one of them to proceed. This causes the exponential in $p$ blowup in our running time at present.

\begin{algorithm}[tb]
   \caption{Sampling from Partition-DPPs}
   \label{alg:partdppalgo}
\begin{algorithmic}
   \STATE {\bfseries Input:} Matrix $A \in \R^{m \times n}$ and $1 \leq k \leq $ rank(A)
   \STATE {\bfseries Output:} Subset $S$ of $k$ rows of $A$ picked with probability proportional to $\det(A_S A_S^T)$
   \STATE Initialize $S \leftarrow \emptyset$ and $B \leftarrow A$.
   \FOR{$t=0$ {\bfseries to} $k-1$}
   \FOR{$i=1$ {\bfseries to} $m$}
   \STATE Compute $p_i = \|b_i\|^2 |c'(C_i C_i^T)|$,
   \\ where $C_i$ is a matrix obtained by projecting each row of $B$ orthogonal to $b_i$ \\
   and $c'(C_i C_i^T)$ is defined as per Theorem \ref{margProbLemma}.
   \ENDFOR

   \STATE Pick $i$ with probability proportional to $p_i$.
   \STATE $S \leftarrow S\cup\{i\}$ and $B\leftarrow C_i$
   \ENDFOR
   \STATE \textbf{return} $S$
\end{algorithmic}
\end{algorithm}

\subsection{MAP Inference for $k$-DPPs}
The maximum a posteriori (MAP) inference problem for $k$-DPPs corresponds to finding a subset $S$ of size $k$ that maximizes $\det{\left(A_{S} A_{S}^{T}\right)}$. This problems is NP-hard to approximate within a multiplicative factor of $2^{\mathcal{O}(k)}$ \cite{CivrilM2013}, a simple greedy algorithm gives a $2^{\mathcal{O}(k \log k)}$-approximation, and the only improvement over the greedy that achieves a $2^{\mathcal{O}(k)}$-approximation is via a convex program that is practically prohibitive \cite{Nikolov15}. Thus, a natural question is whether a greedy initialization can be further improved by some local search heuristic. In this subsection, we give a simple $2^{k \log \min\{k, \kappa(A)\}}$-approximation algorithm this way. Given an $\epsilon > 0$, start with the greedy solution $S$, and pick $i \in S$ and $j \not\in S$ such swapping $i$ and $j$ gives the maximum improvement over the current solution $S$. We do this swap only if the improvement is at least $(1 + \epsilon/k)$ in multiplicative factor. Our experiments show that this gives significantly better results as $k$ grows larger.

\begin{theorem} A greedy initialization followed by $\mathcal{O}\left(k^{2} \log k/ \epsilon\right)$ local search steps as above gives a $2^{k \log \min\{k, \kappa(A)\}}$-approximation to $\underset{\size{S} = k}{\max} \det{\left(A_{S} A_{S}^{T}\right)}$, where $\kappa(A)$ is \small
\[
\kappa(A) = \frac{\lambda_{1} (AA^{T})}{\dfrac{1}{n-k+1} \sum_{t \geq k} \lambda_{t} (AA^{T})}.
\]\normalsize
Note that $\kappa(A)$ is at most the condition number of $AA^{T}$ but in practice could be much better if the ill-conditioning is only due to a small number of bottom eigenvalues. This algorithm takes time linear in the number of non-zero entries of $A$.
\end{theorem}
\begin{proof} See the Appendix. In short, the proof relates $\det(A_{S} A_{S}^{T})$ to the rank-$k$ approximation error for $AA^{T}$, which is why we get $\kappa(A)$ defined as above.
\end{proof}

\section{Experiments}
The effectiveness of DPPs in modeling diversity has been demonstrated in \cite{Kulesza2011,Kulesza2012}. Our experiments focus on comparing Partition-DPP algorithm against the two natural extensions of $k$-DPP to highlight their over/under-representation and other issues in a way that makes them evident to the reader.
\subsection{Limitations of $k$-DPPs and independent $k_{i}$-DPPs}
We study the performance of Partition-DPPs on two real-world image search tasks as follows.
\begin{enumerate}
\item Our first task compares $k$-DPPs, where $k = \sum_{i=1}^{p} k_i$, against DPPs that take partition into account, namely, the independent $k_{i}$-DPPs on each partition and the Partition-DPP. We compare the skew in the number of items per part returned by $k$-DPP to measure its over/under-representation. Note that both the independent $k_i$-DPPs and Partition-DPPs pick exactly $k_{i}$ items from part $i$.
\item Our second task highlights a disadvantage of independent $k_{i}$-DPPs on multi-label images. Suppose each label forms a partition. To ensure disjoint partitions, an image is replicated for its each label and each copy is treated as a separate item. Now independent $k_i$-DPPs may pick the same image multiple times, coming from different partitions, whereas Partition-DPP does not. Due to lack of space, details of this experiment are provided in the appendix.
\end{enumerate}

\subsubsection{Data and Methods}
For the first task, we consider the Yale face database \cite{GeBeKr01,KCLee05} with the cropped images under different lighting conditions. We randomly selected a subset of size 32 out of 38 people under two different poses. For each pose, we selected three lighting conditions - light, medium and dark, which form the partitions. We split the people into four groups of size 8 each. In each group, we define two experiments based on the facial expression. We used the cropped dataset since the images are well-centered, aligned and of the same size. Thus, we create and run 8 different experiments. We computed the Scalable Invariant Feature Transform (SIFT) descriptors \cite{Lowe04} and the corresponding similarity scores between images using the \texttt{vlfeat} toolbox for the first task.

As mentioned above, we compare the Partition-DPP against the $k$-DPP and the independent $k_i$-DPP. For the first task, where we have three partitions, we always selected 6 images. For the second task, we are interested in demonstrating the difference between independent $k_i$-DPPs and Partition-DPPs which stems from the possibility of the former choosing the same element multiple times across different partitions. The 3-tuples considered in this case are $(3,3,3)$ and $(2,3,2)$.

\subsubsection{Results: Yale Face Database Task}
 This task was performed to illustrate that $k$-DPPs can end up giving arbitrarily skewed results. Table 1 shows the various elements that were chosen by the 6-DPP ($k=6$) with the columns separating the elements based on the lighting conditions. People have been labelled from 1 to 8.
\begin{table}[h]
\centering
\begin{tabular}{ | l | c | r |}
  \hline
  Light & Medium & Dark \\ \hline
  2,3,5,6 & 7 & 8 \\
  1,3,4 & 6 & 1,5 \\
  3,5 & 2,4,7 & 6 \\
  4,6,8 & 1 & 2,3 \\
  \hline
\end{tabular}
\begin{tabular}{ | l | c | r |}
  \hline
  Light & Medium & Dark \\ \hline
  4,5,8 & 6,7 & 3 \\
  2,5,7 & 1 & 4,6 \\
  1,4,5 & 3,7 & 8 \\
  1,5 & 2,7 & 3,4 \\
  \hline
\end{tabular}
\label{table:yale6dpp}
\caption{Elements sampled from 6-DPP for the 8 different Yale Task Experiments}
\end{table}
\vspace{-0.2em}
As can be seen, the number of elements in a single partition may be quite high/low or be evenly spread out, although it seems that the SIFT kernel was promoting more well-lit images as compared to the other lighting conditions. This inherent bias of the kernel may be undesirable in presence of partition constraints, e.g. when we want to promote medium or dark images more. Independent $k_i$-DPPs and Partition-DPPs provide the freedom to decide how skewed or uniform the three partitions should be. It should be noted that there were some cases where the same person ended up getting picked due to the SIFT kernel ended up treating the well-lit and dark images of a single person as different. We also show the results obtained by all three methods for the setting of $(k_1, k_2, k_3) = (2,2,2)$ in Table \ref{table:samplingYaleFace}.
\begin{table}[!h]
\begin{tabular}{c|c|c|c} \\ \hline
Method & Light & Medium & Dark\\ \hline
$6$-DPP &
\includegraphics[width=1.25cm,height=1.4cm]{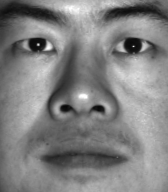}
\includegraphics[width=1.25cm,height=1.4cm]{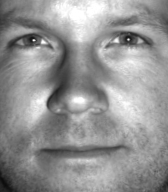}
\includegraphics[width=1.25cm,height=1.4cm]{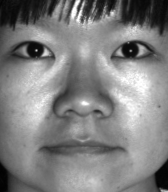}
\includegraphics[width=1.25cm,height=1.4cm]{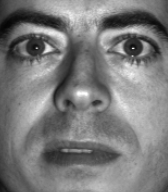} & \includegraphics[width=1.25cm,height=1.4cm]{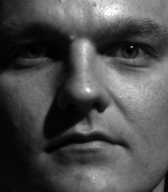} & \includegraphics[width=1.25cm,height=1.4cm]{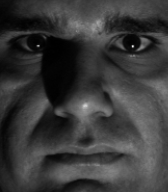}\\\hline
$(2,2,2)$-DPP &\includegraphics[width=1.25cm,height=1.4cm]{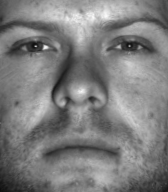}
\includegraphics[width=1.25cm,height=1.4cm]{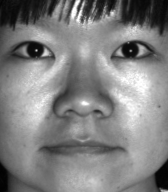}&\includegraphics[width=1.25cm,height=1.4cm]{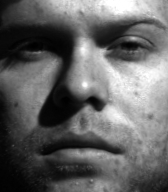}
\includegraphics[width=1.25cm,height=1.4cm]{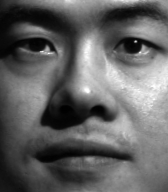}&\includegraphics[width=1.25cm,height=1.4cm]{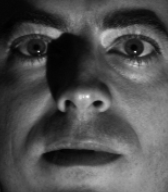}
\includegraphics[width=1.25cm,height=1.4cm]{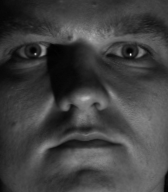}\\ \hline
Partition-DPP & \includegraphics[width=1.25cm,height=1.4cm]{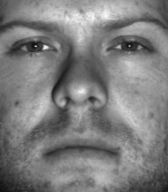}
\includegraphics[width=1.25cm,height=1.4cm]{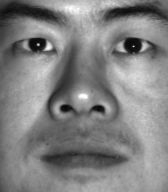} & \includegraphics[width=1.25cm,height=1.4cm]{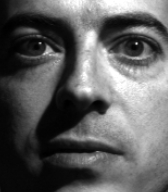}
\includegraphics[width=1.25cm,height=1.4cm]{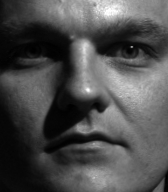} & \includegraphics[width=1.25cm,height=1.4cm]{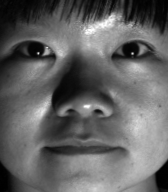}
\includegraphics[width=1.25cm,height=1.4cm]{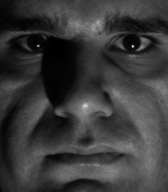} \\ \hline
\end{tabular}
\caption{Sets obtained by sampling from 6-DPP, (2,2,2)-DPP and the Partition-DPP}
\label{table:samplingYaleFace}
\end{table}
\vspace{-0.2cm}

\subsection{Timing Experiments for Partition-DPPs}
While the sampling algorithm for Partition-DPPs is exponential in the number of parts, we present some experiments on synthetic data (since sampling time is independent of kernel and data) with a small number of parts for $k$-DPPs, $k_i$-DPPs and Partition-DPPs. The running time is in seconds. In each table, the first row denotes the number of items in each part $(m_1, \dotsc, m_p)$ and the number of items to be picked from each part $(k_1, \dotsc, k_p)$, respectively. The number of parts $p$ will be clear from context. The next three rows show the method names followed by the running time for that method.
\small
\begin{table}[!htbp]
\begin{tabular}{c||c|c|c|c}\hline
&\makebox[7em]{[15,15], [3,3]}&\makebox[7em]{[15,15], [5,5]}&\makebox[7em]{[24,24], [5,5]}
&\makebox[7em]{[24,24], [10,10]}\\ \hline\hline
$k$-DPP& 0.51&0.85&2.31&5.17\\\hline
$k_i$-DPPs& 0.58&1.01&2.92&5.77\\\hline
Partition-DPP &3.18&7.40&79.45&133.13\\\hline
\end{tabular}

\begin{tabular}{c||c|c|c|c}\hline
&\makebox[7em]{[10,10,10], [2,2,2]}&\makebox[7em]{[10,10,10], [5,5,5]}&\makebox[7em]{[16,16,16], [5,5,5]}
&\makebox[7em]{[6,6,6,6], [1,1,1,1]}\\ \hline\hline
$k$-DPP& 0.51&0.85&2.31&5.17\\\hline
$k_i$-DPPs& 0.58&1.01&2.92&5.77\\\hline
Partition-DPP &3.18&7.40&79.45&133.13\\\hline
\end{tabular}
\caption{Results for timing experiments for Partition-DPPs. Time is in seconds.}
\end{table}
\normalsize
As seen in the table, our algorithm takes some time when there are roughly $50$ elements in total and the number of partitions is $3$ and beyond. For $p=4$, our algorithm takes far too long when $m$ is around $50$. However, given that prior work \cite{GillenwaterMAP2013,Kulesza2011} with DPPs typically has $m$ in the range of $[50, 200]$, our algorithm performs reasonably.
\subsection{MAP Inference for $k$-DPPs}

We compare against the standard Greedy algorithm and the unconstrained Softmax Maximization approach by \cite{GillenwaterMAP2013}. For the number of elements $k$ chosen by the Symmetric Greedy algorithm \cite{BuchbinderFNS12} which gives a $1/3$-approximation ratio for the unconstrained $\log\det$ maximization problem, we compare Greedy and our method as well.

Our first synthetic dataset consists of 50 points in 40 dimensions and each $A_{i,j} \sim 5 * \mbox{Uniform}(0,1)$. To recall, $K=A A^T$. In this case, Symmetric greedy picked 39 elements. The results are shown in Figure 1a. Initially, our algorithm does not give any improvement over the Greedy algorithm but as $k$ grows beyond 20, our algorithm starts giving better results. This is in line with our theoretical approximation ratio as well since initially $k$ is smaller than $\kappa$ but for large $k$, if the growth of $\kappa$ is sublinear, our algorithm starts doing better. We refer the reader to the appendix for exact values for this and another synthetically generated dataset.
\begin{figure}
    \centering
    \begin{subfigure}[b]{0.45\textwidth}
        \includegraphics[width=0.75\textwidth, height=0.6\textwidth]{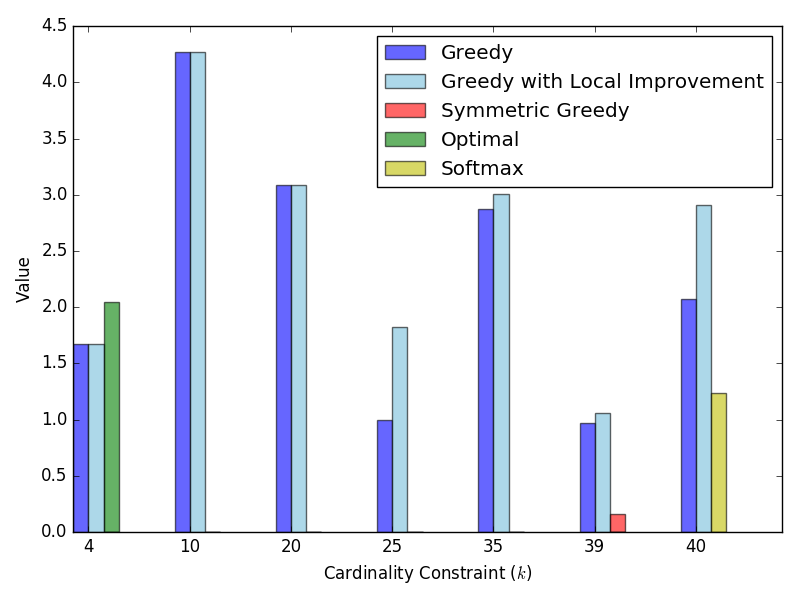}
        \caption{Synthetic Dataset}
        \label{fig:gull}
    \end{subfigure}
    ~ 
    \begin{subfigure}[b]{0.45\textwidth}
        \includegraphics[width=0.75\textwidth, height=0.6\textwidth]{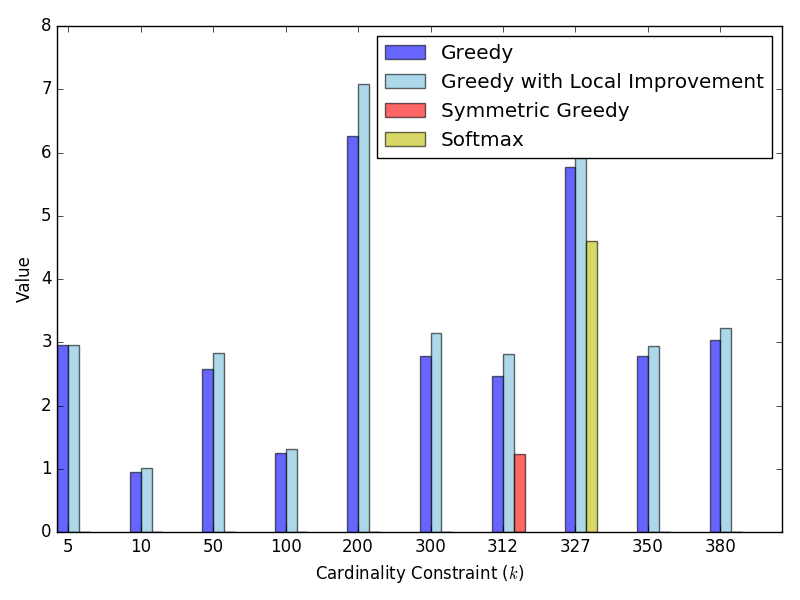}
        \caption{Matched Summarization Dataset}
        \label{fig:tiger}
    \end{subfigure}
    \caption{Results for MAP Inference for $k$-DPPs}
\end{figure}

We also present our results on a real-world dataset used by \cite{GillenwaterMAP2013} for matched summarization. We describe our experiment, which is slightly different than their task of selecting document pairs which have additional similarity measures. Statements of eight main contenders in the 2012 US Republican primary debates are extracted and the same preprocessing as \cite{GillenwaterMAP2013} is performed. Then, a feature matrix $W$ where $W_{qt}$ is the number of times term $t$ appears in quote $q$. The kernel considered is $K=W W^T$. We chose a total of 400 quotes (50 from each candidate). Symmetric Greedy picked 312 elements. The results are presented in Figure 1b.

\normalsize
\vspace{-0.8em}
\section*{Conclusions and Future Work}
\vspace{-0.8em}

We introduce Partition-DPPs for subset selection that respects given partitions or labels that may not be captured by the DPP kernel. Our experiments demonstrate how naive extensions to $k$-DPPs suffer from over/under-representation or duplication, both of are avoided by Partition-DPPs. We quantitatively improve MAP inference for $k$-DPPs on well-conditioned kernels, with experiments supporting our theoretical result. An important direction of future work is improving the sampling time complexity of Partition-DPPs to run in time polynomial in $m, k$ as well as $p$. Another direction is to explore more constrained-DPPs and to provide efficient sampling algorithms for them.
\small
\bibliographystyle{plain}
\bibliography{nips_2016}
\normalsize
\section*{APPENDIX}
\subsection*{Complexity Theoretic Barrier for Exact DPP Sampling under General Matroid Constraints}
 
We show that DPP under transversal matroid constraints can simulate random matchings in bipartite graphs. Transversal matroids are defined using a collection of subsets $S_{1}, S_{2}, \dotsc, S_{k} \subseteq [m]$ that are not necessarily disjoint. A subset $S$ satisfies transversal matroid constraint if $\size{S \cap S_{i}} \leq 1$, for all $1 \leq i \leq k$. See Edmonds and Fulkerson\cite{Edmonds1965} for details.
 
We can define an underlying bipartite graph $(U, V, E)$ where the left vertices are $U = \{S_{1}, S_{2}, \dotsc, S_{k}\}$ and the right vertices $V = \{1, 2, \dotsc, m\}$ and we put an edge between $S_{i}$ and $j$, if $j \in S_{i}$. Suppose we relax the disjointness condition in our partition constraints and use the constraints $\size{S \cap S_{i}} = 1$, for $1 \leq i \leq k$, instead. Then even for $K = I \in \R^{m \times m}$, the corresponding constrained DPP gives a uniform distribution over random perfect matchings in the above bipartite graph. It is well-known that the partition function for this is equivalent to permanent which is \#P-hard to compute by Valiant's theorem \cite{Valiant1979}.
 
Similar to determinantal or Fermion point processes, one can define permanental or Boson point processes. Aaronson-Arkhipov \cite{Aaronson2011} have shown that exact sampling from Boson point processes imply that the polynomial hierarchy collapses to its third level.

\subsection*{Partition-DPPs}
We first provide an example demonstrating that the formula for partition function by \cite{IyerSPP15} does not work for partition matroids. 
Consider a partition $\mathcal{P}_1 = \{1,2\}$ and $\mathcal{P}_2 = \{3,4\}$ of $\{1,2,3,4\}$ with $k_1 = k_2 = 1$ and let $f(S) = \det(K_{S,S})$. The formula in Section 3.6 in \cite{IyerSPP15} gives $Z^1(\emptyset,\mathcal{P}_1) + Z^1(\emptyset,\mathcal{P}_2) = [f(1)+f(2)] + [f(3)+f(4)]$, whereas the partition function for bases of this partition matroid will be $f(\{1,3\})+f(\{1,4\})+f(\{2,3\})+f(\{2,4\})$. This is why sampling from a Partition-DPP is non-trivial and necessitates new algorithms. 

We now present proofs our main results. We will first need a useful theorem known as Schur's identity \cite{SchurIdentiy}.

\begin{lemma} \label{projectionLemma}(Schur's Identity)
Let $A \in \R^{m \times n}$ and $ S,T \subseteq [m], S \cap T = \emptyset$ and $B = A - \pi_S (A)$. Then,
\begin{center}
  $\det(A_{S\cup T} A_{S\cup T}^T) = \det(A_S A_S^T)\det(B_T B_T^T)$
\end{center}
\end{lemma}

We now prove Lemma 3.1 of the main paper.
\begin{proof}(of Lemma 3.1)
  First, it is clear that $c_{0,\ldots,0} = \det(A A^T)$. Next, let $B=A A^T$ and $B' = B - \sum\limits_{i=1}^{p} x_i I_i$ and Perm$(S)$ be the set of permutations of a set $S$. The sign of a permutation $\tau$, denoted by $sgn(\tau)$, is equal to 1 if it is a product of an even number of transpositions and -1 otherwise. 
  
  \begin{center}
    $\det(B - \sum\limits_{i=1}^{p} x_i I_i) = det(B') = \mathlarger{\sum}_{\tau \in \mbox{Perm}([m])} sgn(\tau)B'_{1,\tau(1)}B'_{2,\tau(2)}\ldots B'_{m,\tau(m)}$
  \end{center}
  The term $c_{m_1 - k_1,\ldots,m_p - k_p}x_1^{m_1 - k_1}\ldots x_p^{m_p - k_p}$ is the sum over those permutations $\tau$ which fix some set $S \subseteq [m]$ of size $(m-k)$, where $k = \sum\limits_{l=1}^{p} k_l$ such that for all $ 1 \leq l \leq p, \mbox{ we have } \ |S \cap \PP_l| = m_l - k_l$ and the elements $\prod\limits_{i \in S} B'_{i,i}$ contribute $(-1)^{m-k}x_1^{m_1 - k_1}\ldots x_p^{m_p - k_p}$ and the co-efficient comes from the constant term in 
  
  \begin{center}
    $\mathlarger{\sum}_{\tau \in \mbox{Perm}([m]-S)}\mbox{sgn}(\tau)\mathlarger{\prod}_{i \notin S} B'_{i,\tau(i)}$.
  \end{center}
  The above follows just from the Leibniz formula for the determinant \cite{LeibnizWiki} and then collecting the co-efficients of each monomial.

  Each term in this sum is the $c_0$ term of a principal minor of B satisfying the partition constraints. Thus, the sum is equal to 
  \begin{align*}
    & \mathlarger{\sum}_{\substack{S\subseteq [m] \\ \forall \ i \in [p] \ : \ |S \cap \PP_i| = m_i - k_i}} \det(B_{[m]-S,[m]-S}) \\
    &= \mathlarger{\sum}_{\substack{S\subseteq [m] \\ \forall \ i \in [p] \ : \ |S \cap \PP_i| = k_i}} \det(B_{S,S}) \\
    &= \mathlarger{\sum}_{\substack{S\subseteq [m] \\ \forall \ i \in [p] \ : \ |S \cap \PP_i| = k_i}} \det(A_S A_S^T)
  \end{align*}
  Hence,
  \small
  \begin{align*}
    c_{m_1 - k_1,\ldots,m_p - k_p}  &=  (-1)^{m-k}\mathlarger{\mathlarger{\sum}}_{\substack{S\subseteq [m] \\ \forall \ i \in [p] \ : \ |S \cap \PP_i| = k_i}} \det(A_S A_S^T)
  \end{align*}
\end{proof}

Finally, we prove our main theorem which is Theorem 3.2 in the main paper.

\begin{proof}(of Theorem 3.2)
Define $\mathcal{C}_1$ be a set of constraints as follows 
\begin{center} 
$\mathcal{C}_1 = \bigg\{T \subseteq [m] \ \bigg| \large |T|=k-t-1, \  \forall \ l \in [p]: \ |(S\cup\{i\}\cup T)\cap \PP_l| = k_l \bigg\}$
\end{center} 
In other words, 
\begin{center}
$\mathcal{C}_1 = \bigg\{T \subseteq [m] \ \bigg| \large  |T|=k-t-1, \ |T\cap \PP_j| = k_j - t_j - 1, \ \forall \ l \in [p]\setminus \{j\} : \ |T\cap \PP_l| = k_l-t_l \bigg\} $
\end{center}
We, once again, stress that the above follows since the partitions are considered to be disjoint.
Similarly define $\mathcal{C}_2$ as
\begin{center}
$\mathcal{C}_2 = \bigg\{T \subseteq [m] \ \bigg| \large  |T|=k-t, \ \forall \ l \in [p] : \ |T\cap \PP_l| = k_l-t_l \bigg\} $
\end{center}
Then,
\begin{align*}
  & \mathsf{Pr}(X_{t+1}= i | X_1 = i_1, \ldots, X_t = i_t)  \\
   = & \frac{ \mathlarger{\sum}_{\substack{(i_{t+2},\ldots,i_k) \in [m]^{k-t-1} \\ \forall \ l \in [p] \ : \ |(S\cup \{i\}\cup\{i_{t+2}, \ldots,i_k\}) \cap \PP_l| = k_l}} \mathsf{Pr}({i_1,\ldots,i_t, i,i_{t+2}\ldots,i_k}) }{\mathlarger{\sum}_{\substack{(i_{t+1},\ldots,i_k) \in [m]^{k-t} \\ \forall \ l \in [p] \ : \ |(S\cup\{i_{t+1}, \ldots,i_k\}) \cap \PP_l| = k_l}} \mathsf{Pr}({i_1,\ldots,i_{t},i_{t+1},i_{t+2},\ldots,i_k})} 
   \end{align*}
   \begin{align*}
   = &  \frac{ (k-t-1)! \mathlarger{\sum}_{\mathcal{C}_1} \det(A_{S\cup\{i\}\cup T} A_{S\cup\{i\}\cup T}^T ) }{(k-t)! \mathlarger{\sum}_{\mathcal{C}_2} \det (A_{S\cup T} A_{S\cup T}^T)}  \\
   = & \frac{ \mathlarger{\sum}_{\mathcal{C}_1} \det(A_{S\cup\{i\}\cup T} A_{S\cup\{i\}\cup T}^T ) }{(k-t) \mathlarger{\sum}_{\mathcal{C}_2} \det (A_{S\cup T} A_{S\cup T}^T)}\\
   = & \frac{ \mathlarger{\sum}_{\mathcal{C}_1} \det(A_S A_S^T)\det(B_{\{i\}\cup T} B_{\{i\}\cup T}^T ) }{(k-t) \mathlarger{\sum}_{\mathcal{C}_2} \det(A_S A_S^T) \det (B_{T} B_{ T}^T)} \quad \mbox{by Lemma \ref{projectionLemma}}\\
    = & \frac{ \mathlarger{\sum}_{\mathcal{C}_1} \|b_i\|^2 \det((C_{i})_T (C_{i})_T^T ) }{(k-t) \mathlarger{\sum}_{\mathcal{C}_2} \det (B_{T} B_{T}^T)} \quad \mbox{by Lemma \ref{projectionLemma} applied on B}\\
    = & \frac{ \|b_i\|^2\mathlarger{\sum}_{\mathcal{C}_1}  \det((C_{i})_T (C_{i})_T^T ) }{(k-t) \mathlarger{\sum}_{\mathcal{C}_2} \det (B_{T} B_{T}^T)}\\
    = & \frac{ \|b_i\|^2 |c'(C_i C_i^T)| }{(k-t) |c''(B B^T)|} \quad \mbox{by Lemma 3.1}
\end{align*}
\end{proof}

\subsection*{MAP Inference for $k$-DPPs}
Firstly, denote $d(x,E)$ to be the distance of a vector $x$ to the subspace $E$, i.e., it is the $l_2$-norm of the projection of the $x$ onto the subspace orthogonal to $E$.
\begin{proof}(of Theorem 4.1)
A simple greedy algorithm that starts with $S = \emptyset$ and, in $k$ steps, adds one element at a time that maximizes $\det{\left(A_{S \cup \{i\}} A_{S \cup \{i\}}^{T}\right)}$ and updates $S$ to $S \cup \{i\}$, gives a $2^{\mathcal{O}(k \log k)}$ factor approximation to the MAP inference problem for $k$-DPPs \cite{CivrilM2009}. We can now run a local search where we pick an element $i \in S$ and replace it with some $j \notin S$ that gives maximum improvement as long as
\[
\det{\left(A_{S \setminus \{i\} \cup \{j\}} A_{S \setminus \{i\} \cup \{j\}}^{T}\right)} > \left(1 + \frac{\epsilon}{k}\right) \det{\left(A_{S} A_{S}^{T}\right)},
\]
and update $S$ to $S \setminus \{i\} \cup \{j\}$. Since we start with a greedy initialization that gives $2^{\mathcal{O}(k \log k)}$ approximation, the local improvement must halt in $\mathcal{O}\left(k^{2} \log k/ \epsilon\right)$ steps. When the local improvement stops, for all $i \in S$ and $j \notin S$, we have
\[
\det{\left(A_{S \setminus \{i\} \cup \{j\}} A_{S \setminus \{i\} \cup \{j\}}^{T}\right)} < \left(1 + \frac{\epsilon}{k}\right) \det{\left(A_{S} A_{S}^{T}\right)},
\]
This implies that, for all $i \in S$ and $j \notin S$,
\[
d\left(a_{j}, \operatorname{span}(S \setminus \{i\})\right)^{2} < \left(1 + \frac{\epsilon}{k}\right) d\left(a_{i}, \operatorname{span}(S \setminus \{i\})\right)^{2}.
\]
Therefore, summing the LHS over all $j \notin S$ and $i$ we get
\begin{align*}
& \norm{A - \pi_{S \setminus \{i\}}(A)}_{F}^{2} \\
& \qquad \qquad < \left(1 + \frac{\epsilon}{k}\right) (n-k) d\left(a_{i}, \operatorname{span}(S \setminus \{i\})\right)^{2},
\end{align*}
for all $i \in S$. For any $i \in S$, the LHS is at least the best rank-$(k-1)$ approximation error for $A$ under the Frobenius norm, which is $\sum_{t \geq k} \lambda_{t}(AA^{T})$. Taking the product of the RHS over $i \in S$ and upper bounding $\prod_{i \in S} d\left(a_{i}, \operatorname{span}(S \setminus \{i\})\right)^{2}$ by $\det{(A_{S} A_{S}^{T})}$, we obtain
\[
\frac{1}{\left(1 + \dfrac{\epsilon}{k}\right)^{k}} \left(\frac{1}{n-k}~ \sum_{t \geq k} \lambda_{t} (AA^{T})\right)^{k} < \det{(A_{S} A_{S}^{T})},
\]
Suppose we do not have a $2^{k \log \kappa(A) + 1}$ or $2 \kappa(A)^{k}$-approximation when the local search stops, then $\det{(A_{S} A_{S}^{T})} < \frac{1}{2}~ \kappa(A)^{-k} \underset{\size{S} = k}{\max} \det{(A_{S} A_{S}^{T})} \leq \frac{1}{2}~ \kappa(A)^{-k} \big(\lambda_{1} (AA^{T})\big)^k$, which is at most \small
\[
\frac{1}{2} \left(\frac{1}{n-k}~ \sum_{t \geq k} \lambda_{t} (AA^{T})\right)^{k}, \quad \text{giving a contradiction}.
\]
\end{proof}

\subsection*{Experiments}
\subsubsection*{Corel16k Task for ``Limitations of $k$-DPPs''}
In this section, we describe the data and experimental results of the multi-label image search task which demonstrates another limitation of $k$-DPPs.
\subsection*{Data and DPP Kernel}
For the second task, we use the corel16k dataset \cite{BarnardDFFBJ03}. The images were cropped to consider them to be of the same size and centered. We fix our label space to be a subset of size 3 in two experiments. In the first experiment, the label space is \{sky, sun, tree\}. We took 9 images (numbered 1 to 9) where they have one or more labels from this label set and Table 1 shows the images in each label/partition. Each image had at most 4 labels and those images were chosen which did not have more than 2 labels not from this label set. In the second experiment, the label space considered was \{prop, plane, formation\} and again 9 images were considered. Since the corel16k dataset does not provide the actual images, we only check for duplicates in the set of images sampled. We just used the 46 features provided in the dataset itself. The kernel is just the dot-product of the features of both images.
\begin{table}[h]
\centering
\begin{tabular}{| l | c |}
  \hline
  Label & Corresponding Images \\ \hline
  Sky & 1,4,6,7,8 \\
  Sun & 1,2,3,4,5,6,7 \\
  Tree & 5,6,7,8,9 \\
  \hline
\end{tabular}
\label{table:yale1}
\caption{Labels and Corresponding Images for Experiment 1 of the Corel16k task}
\end{table}
\subsection*{Results}
This task was performed to illustrate the limitation of independent $k_i$-DPPs which is the possible duplication in the selected images. Clearly, $k$-DPPs and Partition-DPPs will not choose the same image multiple times, however, independent $k_i$-DPPs may suffer from this. This is especially likely in cases where some elements, which occur in multiple partitions, are more likely to sampled due to the kernel promoting them. The two tables in Table 3 provide the list of sampled sets (each column denotes a partition) in three different runs of the experiment for a given configuration of $k_i$s. The first three rows in both correspond to the $k_i$ configuration of $(3,3,3)$ and the bottom three correspond to $(2,3,2)$. The images were labelled from 1 to 9. As can be seen from the table, there is considerable duplication of elements in both the cases.

\begin{table}[h]
\centering
\begin{tabular}{ | l | c | r |}
  \hline
  Sky & Sun & Tree \\ \hline
  1,4,7 & 1,3,6 & 5,6,9 \\
  4,7,8 & 1,2,6 & 5,6,7 \\
  1,6,8 & 2,3,7 & 6,8,9 \\
  \hline
  1,4 & 1,3,6 & 6,7 \\
  4,6 & 5,6,7 & 8,9 \\
  7,8 & 2,3,7 & 5,7 \\
  \hline
\end{tabular}
\begin{tabular}{ | l | c | r |}
  \hline
  Prop & Plane & Formation \\ \hline
  1,3,4 & 4,6,9 & 5,6,9 \\
  2,3,6 & 2,4,7 & 4,5,8 \\
  4,6,9 & 6,8,9 & 4,5,8 \\
  \hline
  1,2 & 2,7,9 & 4,8 \\
  4,9 & 3,4,8 & 5,6 \\
  1,3 & 2,3,9 & 6,9 \\
  \hline
\end{tabular}
\label{table:kvecdpp}
\caption{Elements sampled from the two different sets of the independent $k_i$-DPPs to illustrate duplication in chosen elements}
\end{table}

\subsection*{MAP Inference for $k$-DPPs}
We first present the results of the first synthetic task which consists of 50 points in 40 dimensions and each $A_{i,j} \sim 5 * $Uniform(0,1). The results are provided in Table 3.
\begin{table}[!h]
\centering
\begin{tabular}{|c|@{}c| @{}c|}\hline
$k$ & Method & Value \\\hline
$k=4$
&
\begin{tabular}{c}
Greedy  \\
Our method \\
Optimal \\
\end{tabular}
&
\begin{tabular}{c}
 1.67e9 \\
1.67e9\\
2.05e9 \\
\end{tabular}
\tabularnewline\hline
$k=10$
&
\begin{tabular}{c}
Greedy  \\
Our method \\
\end{tabular}
&
\begin{tabular}{c}
 4.27e16 \\
4.27e16\\
\end{tabular}
\tabularnewline\hline
$k=20$
&
\begin{tabular}{c}
Greedy  \\
Our method \\
\end{tabular}
&
\begin{tabular}{c}
 3.09e31 \\
3.09e31\\
\end{tabular}
\tabularnewline\hline
$k=25$
&
\begin{tabular}{c}
Greedy  \\
Our method \\
\end{tabular}
&
\begin{tabular}{c}
 9.93e37 \\
1.82e38\\
\end{tabular}
\tabularnewline\hline
\end{tabular}
\quad
\begin{tabular}{|c|@{}c| @{}c|}\hline
$k$ & Method & Value \\\hline

$k=35$
&
\begin{tabular}{c}
Greedy  \\
Our method \\
\end{tabular}
&
\begin{tabular}{c}
 2.87e48 \\
3.01e48\\
\end{tabular}
\tabularnewline\hline
$k=39$
&
\begin{tabular}{c}
Greedy  \\
Our method \\
Symmetric Greedy \\
\end{tabular}
&
\begin{tabular}{c}
 9.69e50 \\
1.06e51\\
1.59e50\\
\end{tabular}
\tabularnewline\hline
$k=40= $ rank
&
\begin{tabular}{c}
Greedy  \\
Our method \\
Softmax\\
\end{tabular}
&
\begin{tabular}{c}
 2.07e51 \\
2.91e51\\
1.24e51\\
\end{tabular}
\tabularnewline\hline
\end{tabular}
\caption{Results of the first synthetic dataset for MAP Inference}
\end{table}

Our second synthetic dataset consists of 200 points in 200 dimensions and each $A_{i,j} \sim \mbox{Uniform}(0,1)$. In this case, Greedy, our algorithm and Symmetric Greedy all picked all 200 of the elements. The results are shown in Table 4.
\small
\begin{table}[!h]
\centering
\begin{tabular}{|c|@{}c| @{}c|}\hline
$k$ & Method & Value \\\hline
$k=5$
&
\begin{tabular}{c}
Greedy  \\
Our method \\
\end{tabular}
&
\begin{tabular}{c}
 3.62e7 \\
3.89e7\\
\end{tabular}
\tabularnewline\hline
$k=10$
&
\begin{tabular}{c}
Greedy  \\
Our method \\
\end{tabular}
&
\begin{tabular}{c}
 1.30e14 \\
1.48e14\\
\end{tabular}
\tabularnewline\hline
$k=20$
&
\begin{tabular}{c}
Greedy  \\
Our method \\
\end{tabular}
&
\begin{tabular}{c}
 5.34e26 \\
5.99e26\\
\end{tabular}
\tabularnewline\hline
\end{tabular}
\quad
\begin{tabular}{|c|@{}c| @{}c|}\hline
$k$ & Method & Value \\\hline
$k=50$
&
\begin{tabular}{c}
Greedy  \\
Our method \\
\end{tabular}
&
\begin{tabular}{c}
 2.96e62 \\
3.18e62\\
\end{tabular}
\tabularnewline\hline
$k=100$
&
\begin{tabular}{c}
Greedy  \\
Our method \\
\end{tabular}
&
\begin{tabular}{c}
 5.14e98 \\
5.61e98\\
\end{tabular}
\tabularnewline\hline
$k=150$
&
\begin{tabular}{c}
Greedy  \\
Our method \\
\end{tabular}
&
\begin{tabular}{c}
 1.03e155 \\
1.08e155\\
\end{tabular}
\tabularnewline\hline
\end{tabular}
\caption{Results of the second synthetic dataset for MAP Inference}

\end{table}
\normalsize

For this dataset, the kernel is very well-conditioned and our algorithm beats greedy very early on.

We finally have the results for the matched summarization real-world task of \cite{GillenwaterMAP2013} in Table 5.
\begin{table}[!h]
\centering
\begin{tabular}{|c|@{}c| @{}c|}\hline
$k$ & Method & Value \\\hline
$k=5$
&
\begin{tabular}{c}
Greedy  \\
Our method \\
\end{tabular}
&
\begin{tabular}{c}
 2.95e3 \\
2.95e3\\
\end{tabular}
\tabularnewline\hline
$k=10$
&
\begin{tabular}{c}
Greedy  \\
Our method \\
\end{tabular}
&
\begin{tabular}{c}
 9.55e5 \\
1.01e6\\
\end{tabular}
\tabularnewline\hline
$k=25$
&
\begin{tabular}{c}
Greedy  \\
Our method \\
\end{tabular}
&
\begin{tabular}{c}
 8.07e12 \\
8.54e12\\
\end{tabular}
\tabularnewline\hline
$k=50$
&
\begin{tabular}{c}
Greedy  \\
Our method \\
\end{tabular}
&
\begin{tabular}{c}
 2.58e23 \\
2.83e23\\
\end{tabular}
\tabularnewline\hline
$k=100$
&
\begin{tabular}{c}
Greedy  \\
Our method \\
\end{tabular}
&
\begin{tabular}{c}
 1.25e42 \\
1.32e42\\
\end{tabular}
\tabularnewline\hline
$k=150$
&
\begin{tabular}{c}
Greedy  \\
Our method \\
\end{tabular}
&
\begin{tabular}{c}
 4.77e57 \\
4.89e57\\
\end{tabular}
\tabularnewline\hline
$k=200$
&
\begin{tabular}{c}
Greedy  \\
Our method \\
\end{tabular}
&
\begin{tabular}{c}
 6.26e69 \\
7.09e69\\
\end{tabular}
\tabularnewline\hline
\end{tabular}
\quad
\begin{tabular}{|c|@{}c| @{}c|}\hline
$k$ & Method & Value \\\hline

$k=250$
&
\begin{tabular}{c}
Greedy  \\
Our method \\
\end{tabular}
&
\begin{tabular}{c}
 1.03e78 \\
1.11e78\\
\end{tabular}
\tabularnewline\hline
$k=300$
&
\begin{tabular}{c}
Greedy  \\
Our method \\
\end{tabular}
&
\begin{tabular}{c}
 2.79e81 \\
3.15e81\\
\end{tabular}
\tabularnewline\hline
$k=312$
&
\begin{tabular}{c}
Greedy  \\
Symmetric Greedy \\
Our method \\
\end{tabular}
&
\begin{tabular}{c}
 2.46e81 \\
 6.63e80\\
2.82e81\\
\end{tabular}
\tabularnewline\hline
$k=327$
&
\begin{tabular}{c}
Greedy  \\
Softmax \\
Our method \\
\end{tabular}
&
\begin{tabular}{c}
 5.78e80 \\
 6.05e80\\
4.63e80\\
\end{tabular}
\tabularnewline\hline
$k=350$
&
\begin{tabular}{c}
Greedy  \\
Our method \\
\end{tabular}
&
\begin{tabular}{c}
 2.78e78 \\
2.94e78\\
\end{tabular}
\tabularnewline\hline
$k=380$
&
\begin{tabular}{c}
Greedy  \\
Our method \\
\end{tabular}
&
\begin{tabular}{c}
 3.03e72 \\
3.22e72\\
\end{tabular}
\tabularnewline\hline
\end{tabular}
\caption{Results of the real-world dataset for MAP Inference}
\end{table}


\end{document}